\documentclass[aip,reprint,amsmath,amssymb]{revtex4-1}

\usepackage{graphicx}
\usepackage{dcolumn}
\usepackage{float}
\usepackage{bm}
\usepackage{amssymb}
\usepackage{lipsum}

\newcommand*{\citen}[1]{%
  \begingroup
    \romannumeral-`\x 
    \setcitestyle{numbers}%
    \cite{#1}%
  \endgroup
}

\begin{document}

\title{Flatness and Intrinsic Curvature of Linked-Ring Membranes}

\author{James M. Polson}
\affiliation{ Department of Physics, University of Prince Edward Island,
550 University Avenue, Charlottetown, Prince Edward Island, C1A 4P3, Canada }
\author{Edgar J. Garcia}
\affiliation{ Department of Physics and Astronomy, California State University, 
Long Beach, California, 90840, USA}
\author{Alexander R. Klotz}
\affiliation{ Department of Physics and Astronomy, California State University,
Long Beach, California, 90840, USA}

\date{\today}

\begin{abstract}
{Recent experiments have elucidated the physical properties of kinetoplasts, which are chain-mail-like structures found in the mitochondria of trypanosome parasites formed from catenated DNA rings. Inspired by these studies, we use Monte Carlo simulations to examine the behavior of two-dimensional networks (``membranes'') of linked rings. For simplicity, we consider only identical rings that are circular and rigid and that form networks with a regular linking structure. We find that the scaling of the eigenvalues of the shape tensor with membrane size are consistent with the behavior of the flat phase observed in self-avoiding covalent membranes. Increasing ring thickness tends to swell the membrane. Remarkably, unlike covalent membranes, the linked-ring membranes tend to form concave structures with an intrinsic curvature of entropic origin associated with local excluded-volume interactions. The degree of concavity increases with increasing ring thickness and is also affected by the type of linking network. The relevance of the properties of linked-ring model membranes to those observed in kinetoplasts is discussed.}
\end{abstract}

\maketitle

\section{Introduction}

Recent interest in topologically complex soft materials has lead to the fabrication and characterization of molecules with non-covalent connectivity, including molecular knots and linked-ring networks known as catenanes.\cite{hart2021material} Molecular catenanes can be created synthetically with techniques such as metallo-organic complexation, \cite{wu2017poly} but several forms of catenated macromolecules or ``Olympic gels''\cite{deGennes_book,raphael1997progressive} are known to form naturally. These include the HK97 virus that has a capsid made of catenated proteins,\cite{zhou2015protein} and catenated DNA molecules that occur in the mitochondria of cancerous cells.\cite{clayton1967circular} The most extreme example of ``molecular chainmail'' is the kinetoplast. A kinetoplast is a complex DNA structure found in the mitochondria of trypanosome parasites, consisting of thousands of circular DNA molecules, known as minicircles, topologically linked in a two-dimensional network.\cite{englund2014passion} Kinetoplasts are part of an RNA editing mechanism that allows mutated metabolic genes to be expressed, and the network of minicircles is believed to have a honeycomb topology with an average ``valence'' of 3. \cite{chen1995topology}
 
Recently, kinetoplasts have been investigated as a model experimental system for studying the physics of 2D polymers and catenated materials. \cite{klotz2020equilibrium} It was observed that kinetoplasts from the \textit{Crithidia fasciculata} parasite in free solution exhibit the behavior expected of a thermalized elastic membrane but also have a strong intrinsic curvature. Subsequently, Soh et al. showed that stretched kinetoplasts form a metastable deformed state \cite{soh2020deformation} and undergo isotropic changes in size when the buffer ionic conditions are varied.\cite{soh2020ionic} One salient question is the degree to which their exotic catenated structure affects their physical properties, as distinct from their two-dimensional network topology: to what extent is a catenated membrane different from a covalent or tethered membrane? The lengthscale of minicircle catenation is typically below the lengthscale of optical microscopy and the deformations achieved through microfluidic stretching\cite{soh2020ionic} or confinement\cite{klotz2020equilibrium, soh2021equilibrium} may not be sufficient to distinguish their effects.

One predicted feature of self-avoiding two-dimensional polymers is their ``flatness,'' referring to the fact that the length-scale of their surface undulations is predicted to have a weaker dependence on molecular weight than the in-plane dimensions of the polymer, leading to an effectively infinite persistence length independent of molecular composition. While incorporating a local energetic bending rigidity into a model is sufficient to induce flatness, an effective bending rigidity of entropic origin can also arise solely through excluded-volume interactions. As noted by {Kantor} and Kremer, \cite{kantor1993excluded} local excluded-volume interactions provide the bending rigidity that leads to flatness, but then play no further role at larger distances. Flatness has been identified for various models of self-avoiding membranes in numerous simulations.\cite{plischke1988absence,abraham1989molecular,grest1991self,zhang1996molecular,popova2007structure,popova2008anomalous,mizuochi2014dynamical} While kinetoplasts have the appearance of a smooth but curved open membrane, it is not known from experiments whether catenated 2D materials exhibit the predicted flat phases of 2D polymers.

The apparent curvature of kinetoplasts in solution is not known to be an essential part of the trypanosome gene editing apparatus, nor is it known whether it arises due to ``purse-string'' effects at the edge,\cite{barker1980ultrastructure} the topology of the network or defects therein, or a subtler entropic mechanism. There has been recent interest in the spontaneous curvature of thermalized planar materials due to lattice impurities\cite{plummer2020buckling} as well as the influence of intrinsic curvature on defect dynamics,\cite{agarwal2020simple} but the spontaneous equilibrium curvature of membrane-like polymers without explicit curvature has not been investigated. Recently, Soh and Doyle showed that the apparent curvature of kinetoplasts vanishes when they are strongly confined.\cite{soh2021equilibrium}

In this study, we use Monte Carlo simulations to investigate the equilibrium statistical properties 
of catenated membranes.  {While other recent simulation studies have examined the statistical and dynamical 
properties of similar systems, including catenane dimers,\cite{amici2019topologically, 
caraglio2017mechanical} poly-catenanes,\cite{rauscher2018topological, wu2019mechanical, rauscher2020dynamics, 
rauscher2020thermodynamics, dehaghani2020effects} Olympic gels,\cite{lang2014swelling, fischer2015formation, 
lang2015olympic} as well as the linking statistics of ring polymers under confinement,\cite{michieletto2015kinetoplast, 
dadamo2017linking, diao2015orientation} this is the first simulation study of a catenated membrane, to
our knowledge. The model membrane consists of identical rigid circular rings connected in 2D lattices.  
Although the simulations in 
Refs.~\citen{amici2019topologically, caraglio2017mechanical, rauscher2018topological, 
wu2019mechanical, rauscher2020dynamics, rauscher2020thermodynamics, dehaghani2020effects, lang2014swelling, 
fischer2015formation, lang2015olympic, michieletto2015kinetoplast, dadamo2017linking, diao2015orientation}
use flexible-chain models, we find the use of rigid rings to be a necessary simplification for
computational efficiency.}
We are interested in the growth of out-of-plane 
fluctuations and spontaneous curvature with respect to the number of rings in the network, the extent 
to which these features deviate from those found in covalently-connected membranes, as well as the 
effects of ring thickness, lattice shape and linking topology. As observed in the case for covalent 
membranes, we find that linked-ring membranes have a flat topology. Remarkably, we also find that they 
form concave structures qualitatively similar to those observed in kinetoplasts.

\section{Model and Methods}

\begin{figure}[!ht]
    \centering
    \includegraphics[width=\columnwidth]{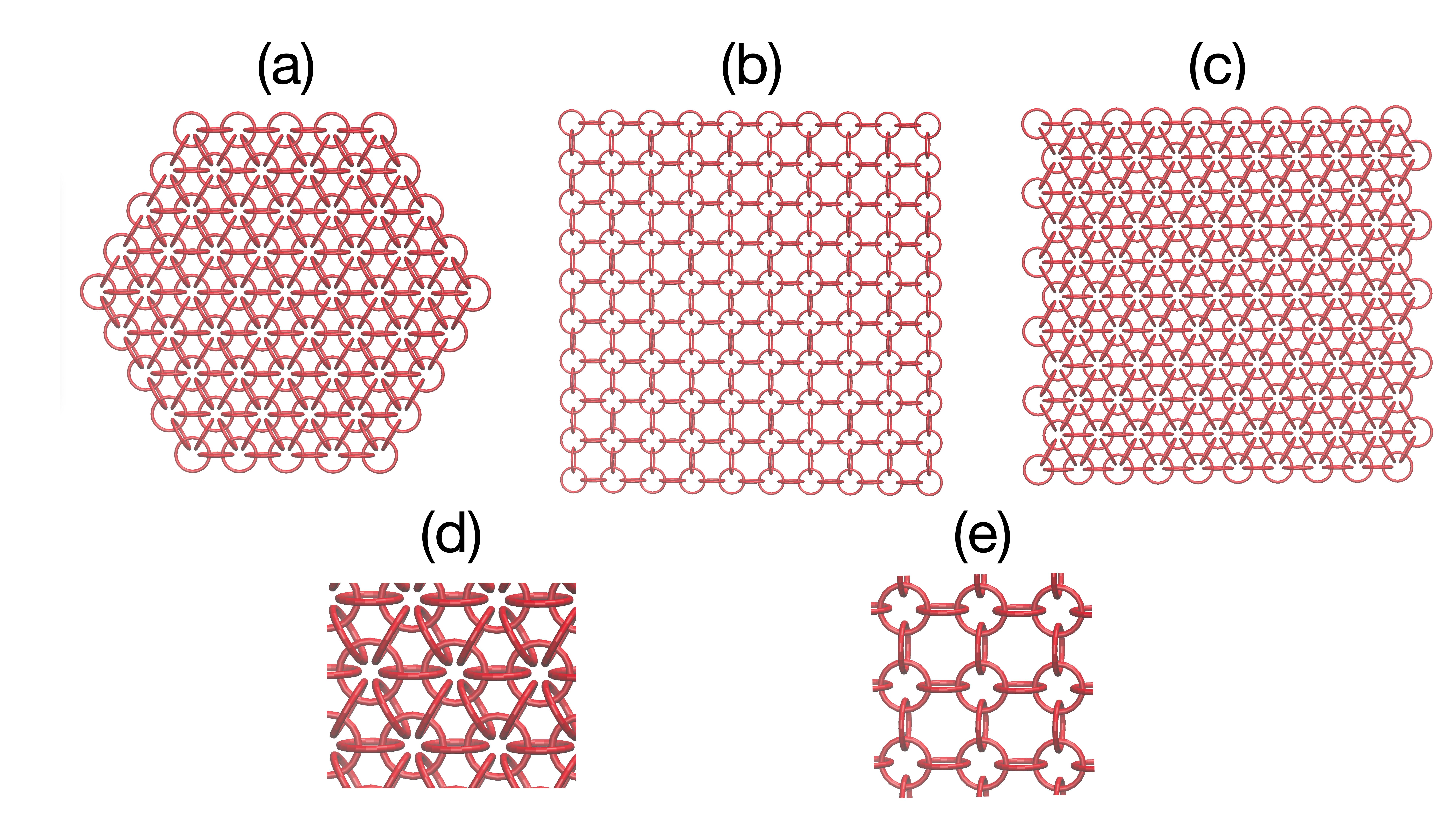}
    \caption{Snapshots of linked-ring membranes illustrating the shapes and linking topologies examined in this study. (a) Hexagonal-shaped membrane with triangular-lattice linking (HT) and a size of $M$=9. (b) Square-shaped membrane with square-lattice linking (SS) and a size of $M$=10. (c) Approximately square membrane with triangular lattice linking (ST) with dimensions $M_1$=10 and $M_2$=11. (d) Close-up illustration of linking structure for HT and ST membranes. (e) Close-up illustration of linking structure for the SS membrane.}
    \label{fig:illust}
\end{figure}

We use Monte Carlo (MC) simulations to study membranes composed of interlocking rigid circular rings. Figure~\ref{fig:illust} illustrates the three membrane structures examined in this study. The membrane of Fig.~\ref{fig:illust}(a) has a hexagonal shape and triangular lattice structure (HT).  It is composed of two types of rings: those with a linking valence of 6 (except at the edges where the valence is either 3 or 4), and those with a valence of 2. This membrane resembles tethered membranes used in previous simulations studies,\cite{kantor1986statistical,abraham1989molecular,grest1991self,popova2007structure,popova2008anomalous,mizuochi2014dynamical} with the 6-valence rings analogous to vertices or particles and the 2-valence rings analogous to the connecting bonds. The membrane size $M$ is defined as the number of 6-valence rings that span the structure from one corner through the center to the opposite corner. As an illustration, Figure~\ref{fig:illust}(a) shows a membrane with $M$=9. Figure~\ref{fig:illust}(b) shows a square membrane with a square-lattice structure (SS) composed of rings with a linking valence of 4 (except at edges and corners) as well as those with a valence of 2. For this linking topology, we consider only membranes that are square when stretched out, as illustrated in the figure. The membrane size $M$ is the number of 4-valence rings along the side of the square. As an illustration, $M$=10 for the membrane in the figure. Figure~\ref{fig:illust}(c) shows the third membrane type examined: a membrane with triangular-lattice linking, as for HT membranes, but whose shape is (approximately) square, as for SS membranes. We call these ST membranes. This model is employed in some calculations to determine whether any observed differences in behavior between HT and SS membranes are caused by the linking topology or the membrane shape. Integer lattice sizes for the 6-valence rings $M_1$ and $M_2$ are chosen to best approximate a square shape. As an illustration, $M_1$=10 and $M_2$=11 for the membrane in Fig.~\ref{fig:illust}(c).

We examine membranes with rings of finite and zero thickness. The diameter of the rings, $D$, is chosen to be the unit of length, i.e. $D$=1. Rings of finite thickness have a circular cross section when bisected by a plane containing the ring normal. The diameter of this cross section, $w$ defines the thickness of the ring. For $w>0$, the diameter is the distance between the centers of the two circular cross sections. We consider a ring thickness in the range $w=0-0.2D$.

The MC simulations use the standard Metropolis methodology. For convenience, the initial positions and orientations of the rings are chosen to  correspond the structures shown in Fig.~\ref{fig:illust}. Two types of MC trial moves are carried out for randomly selected rings: random displacement and random rotation about a randomly chosen axis. Trial moves are accepted or rejected based on whether they preserve the original linking structure, i.e., rings that are originally linked must stay linked, and rings that were originally unlinked must remain so. Any move that violates these constraints is rejected. For rings with $w>0$ moves are also rejected if the volumes occupied by the rings overlap, as determined using the method described in Ref.~\citen{vranek2002fast}. (A detailed description of the algorithms used for testing for interlocking rings and overlap of finite-$w$ rings is presented in the ESI.\dag) Moves that preserve the link structure and do not result in such overlap are accepted. Maximum displacement and rotation angles are chosen to yield an acceptance ratio in the range 30 -- 50\%. Displacement and rotation moves are selected with equal probability. A MC cycle is defined as a sequence of $N$ consecutive trial moves, where $N$ is the total number of rings in the membrane. Thus, during each cycle {an attempt is made to either translate or rotate each ring once,} on average. Prior to data sampling, the system is equilibrated for a period chosen to ensure the complete decay of the initial transients in the histories of all measured quantities. Equilibration periods were typically ${\cal O}(10^6)$ MC cycles, and production runs were in the range of $3-5\times 10^6$ MC cycles in duration. Typically, the results of between 10 and 200 independent simulations were averaged to achieve reasonable statistical accuracy, with larger systems requiring more averaging.

\begin{figure}[!ht]
    \centering
    \includegraphics[width=0.65\columnwidth]{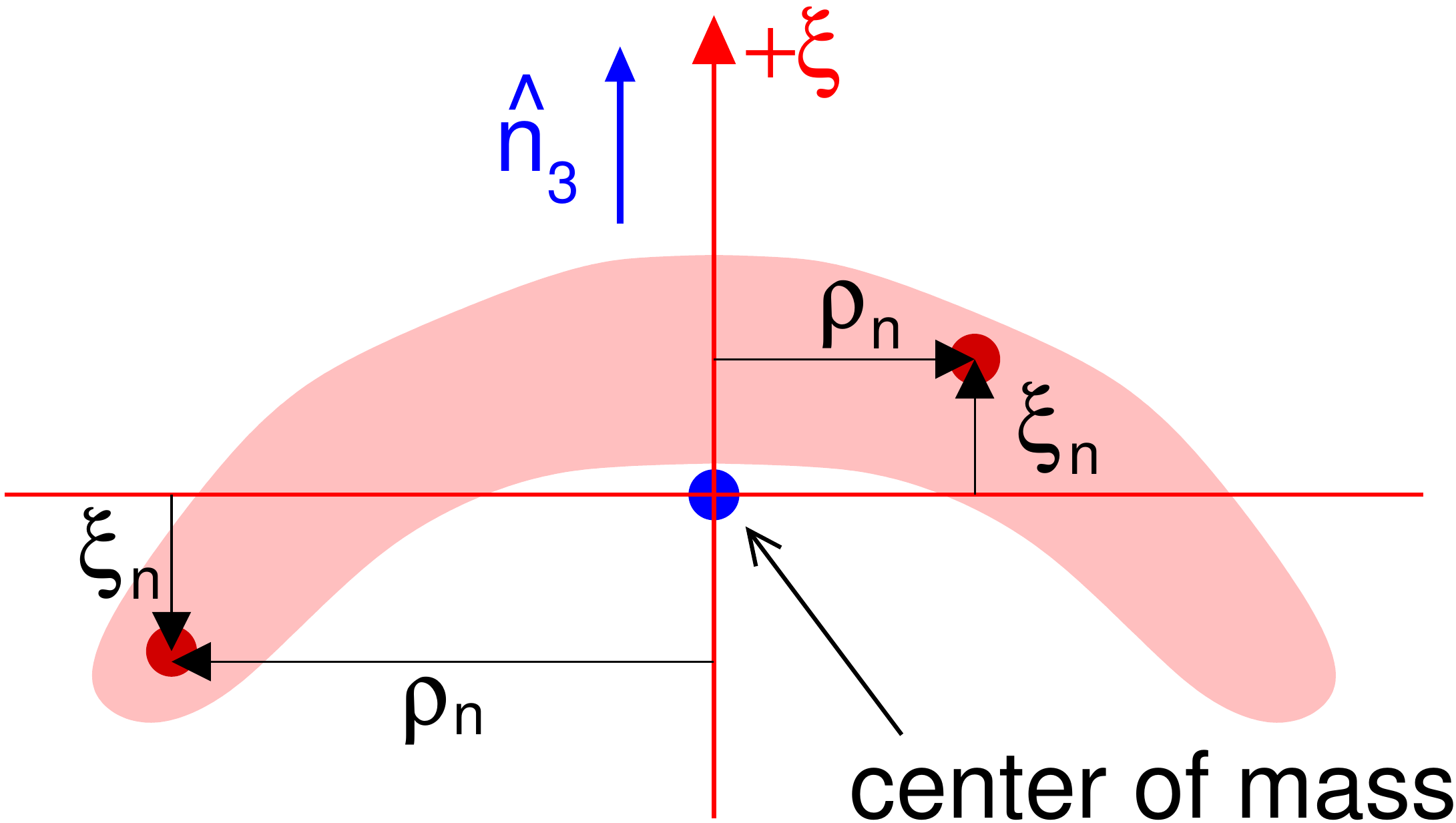}
    \caption{Illustration of quantities used in the definition of concavity $\zeta$ in Eq.~(\ref{eq:concavity}). $\xi_n$ and $\rho_n$ are the position coordinates of ring $n$ in a coordinate system with $\xi$ aligned along $\hat{n}_3$, the eigenvector for the smallest shape-tensor eigenvalue that passes through the membrane center of mass. The shaded region represents a cross section of the membrane in the plane containing $\hat{n}_3$ and the center of mass. {Note that the figure illustrates the case of a membrane with $\zeta<0$.} }
    \label{fig:zeta_illust}
\end{figure}

The principal quantity of interest in this study is the shape tensor, whose components, $S_{\alpha\beta}$, are defined
\begin{eqnarray}
S_{\alpha\beta} = \frac{1}{N} \sum_{i=1}^N (r_{i,\alpha}-r_\alpha^{\rm CM})
(r_{i,\beta}-r_\beta^{\rm CM}),
\label{eq:shape}
\end{eqnarray}
where $r_{i,\alpha}$ is the $\alpha$-coordinate of the position of the center of the $i$th ring, and where $r_\alpha^{\rm CM}$ is likewise the $\alpha$-coordinate for the center of mass of the membrane. The instantaneous {eigenvalues} are denoted $R_1^2$, $R_2^2$, and $R_3^2$, where we choose $R_1^2\geq R_2^2\geq R_3^2$. The corresponding eigenvectors are denoted $\hat{n}_1$, $\hat{n}_2$ and $\hat{n}_3$. Note that the radius of gyration is related by $R_{\rm g}^2 = R_1^2+R_2^2+R_3^2$. We also introduce a measure of membrane concavity, $\zeta$ as follows:
\begin{eqnarray}
\zeta = \frac{1}{N}\sum_{n=1}^N \xi_n \rho_n
\label{eq:concavity}
\end{eqnarray}
Here, $\xi_n$ is a coordinate of the position of the $n$th ring measured along the $\xi$ axis, which is aligned with $\hat{n}_3$. The $\xi$ axis is also defined to pass through the center of mass, which defines the point where $\xi=0$. In addition, $\rho_n$ is the transverse distance of ring $n$ to the nearest point on the $\xi$ axis. Note that $\zeta$ tends to be appreciably non-zero when the rings close to the membrane center are on one side of the center of mass and rings further from the center are on the other side, i.e., where the membrane has a concave structure. The quantities used to define $\zeta$ are illustrated in Figure~\ref{fig:zeta_illust}. {Note that correctly distinguishing between positive and negative values of $\zeta$ requires resolving the ambiguity in choosing between two possible directions of $\hat{n}_3$. This is done by exploiting the fixed connectivity of the membrane rings. The details are discussed in the ESI.\dag}

Of particular interest is the concavity probability distribution, ${\cal P}(\zeta)$, and the related free-energy function, defined by $F(\zeta)/k_{\rm B}T=-\ln {\cal }{\cal P}(\zeta)$, where $k_{\rm B}$ is Boltzmann's constant and $T$ is absolute temperature. In some systems the distributions obtained from simple sampling over 100--200 simulations are averaged to obtain reliable estimates of $F(\zeta)$ over the range of interest for $\zeta$. In other systems a sizeable free energy barrier precludes accurate estimates of $F$ in the barrier region. In such cases, we employ a multiple-histogram method that incorporates umbrella sampling.\cite{frenkel2002understanding} The method was used for the case where the barrier heights exceed approximately $3k_{\rm B}T$. As in previous studies where one of us has employed this method (see, e.g., Ref.~\citen{polson2013simulation}), we refer to the method as the Self-Consistent Histogram (SCH) method. To implement the SCH method, we carry out many independent simulations, each of which employs a unique ``window potential'' of the form:
\begin{eqnarray}
{W_i(\zeta)}=\begin{cases} \infty, \hspace{8mm} \zeta<\zeta_i^{\rm min} \cr 0,
\hspace{1cm} \zeta_i^{\rm min}<\zeta<\zeta_i^{\rm max} \cr \infty,
\hspace{8mm} \zeta>\zeta_i^{\rm max} \cr
\end{cases}
\label{eq:winPot}
\end{eqnarray}
where $\zeta_i^{\rm min}$ and $\zeta_i^{\rm max}$ are the limits that define the range of $\zeta$ for the $i$-th window.  Within each window of $\zeta$, a probability distribution $p_i(\zeta)$ is calculated in the simulation. The window potential width, $\Delta \zeta \equiv \zeta_i^{\rm max} - \zeta_i^{\rm min}$, is chosen to be sufficiently small that the variation in $F$ does not exceed $\approx 2k_{\rm B}T$. The windows are chosen to overlap with half of the adjacent window, such that $\zeta^{\rm max}_{i} = \zeta^{\rm min}_{i+2}$.  The window width was typically in the range $\Delta \zeta = 0.1D-0.2D$. The SCH algorithm was employed to reconstruct the unbiased distribution, ${\cal P}(\zeta)$, from the $p_i(\zeta)$ histograms. The free energy follows from the relation $F(\zeta) = -k_{\rm B}T\ln {\cal P}(\zeta)+{\rm const}$. We choose the constant such that $F=0$ at $\zeta=\zeta_{\rm min}$, where $\zeta_{\rm min}$ is the location of the free-energy minimum.

A derivation of the histogram reconstruction method is described in Ref.~\citen{frenkel2002understanding}. A detailed description of applying the methodology to measure the polymer translocation free-energy function is described in detail in Ref.~\citen{polson2013simulation}.

In the results presented below, distances are measured in units of the ring diameter, $D$, and free energy is measured in units of $k_{\rm B}T$.

\section{Results and Discussion}

\begin{figure}[!ht]
    \centering
    \includegraphics[width=\columnwidth]{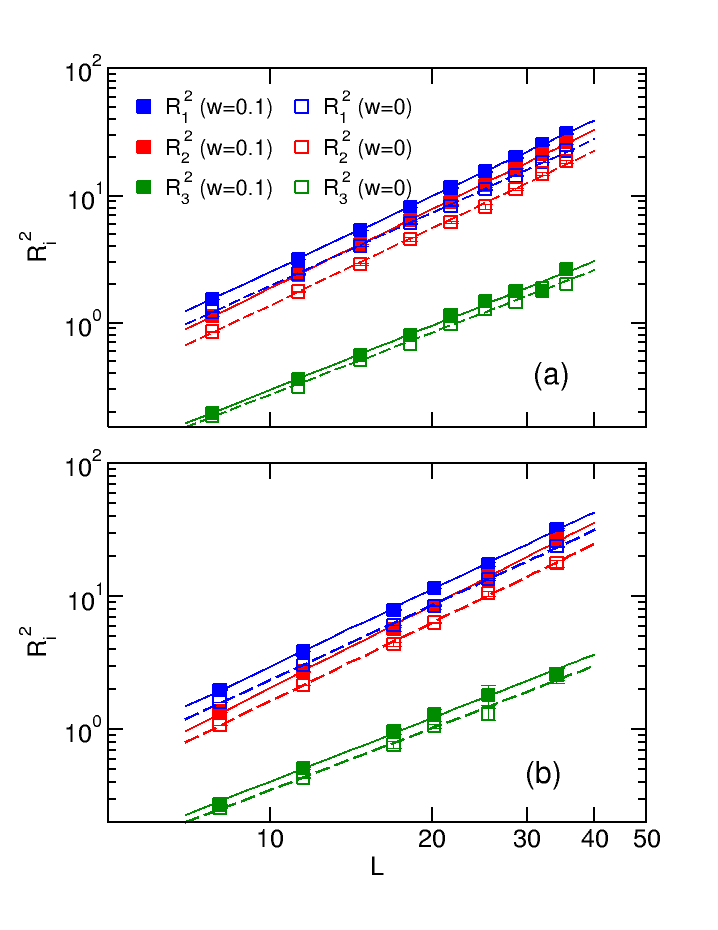}
    \caption{Scaling of eigenvalues of the shape tensor as a function of membrane size, $L\equiv\sqrt{N}$, where $N$ is the total number of rings in the membrane. (a) Results for HT membranes for ring thickness of $w$=0 and $w$=0.1. (b) Results for SS membranes for ring thickness of $w$=0 and $w$=0.1.}
    \label{fig:Rscale}
\end{figure}

Figure~\ref{fig:Rscale} shows the scaling of the shape-tensor eigenvalues, $R_1^2$, $R_2^2$ and $R_3^2$ with respect to $L\equiv\sqrt{N}$, where $N$ is the total number of rings in the membrane. Since $N$ is approximately proportional to the average surface area of the membrane, $L$ is a rough measure of its span measured along the surface. Figure~\ref{fig:Rscale}(a) shows the scaling results for the HT membrane depicted in Fig.~\ref{fig:illust}(a), while Fig.~\ref{fig:Rscale}(b) shows results for the SS membrane illustrated in Fig.~\ref{fig:illust}(b). In each case, results for ring thickness of $w$=0 and $w$=0.1 are shown. In all cases, the eigenvalues exhibit power-law scaling. The solid and dashed lines are the best-fit curves for $R_i^2\propto L^{2\nu_i}$. The values of the scaling exponents $\nu_i$ are presented in Table~\ref{tab:1}.

\begin{table}[!htp]
\begin{center}
\begin{tabular}[t]{c    c     c     c     c}
\hline\hline
           & \multicolumn{2}{c}{HT} & \multicolumn{2}{c}{SS} \\
\hline
           & $w$=0 & $w$=0.1 & $w$=0 & $w$=0.1 \\
           \hline
~$\nu_1$ ~~ & ~0.95$\pm$0.01 ~ & ~ 0.99$\pm$ 0.01~ & ~0.92$\pm$0.02 ~ & ~0.97$\pm$0.02~\\
~$\nu_2$ ~~ & ~0.99$\pm$0.01 ~ & ~ 1.03$\pm$ 0.01~ & ~0.97$\pm$0.01 ~ & ~1.02$\pm$0.03~\\
~$\nu_3$ ~~ & ~0.79$\pm$0.02 ~ & ~ 0.84$\pm$ 0.01  & ~0.73$\pm$0.02 ~ & ~0.84$\pm$0.03~\\
\hline \hline
\end{tabular}
\end{center}
\caption{Scaling exponents, $\nu_i$, for the power-law fits to the data shown in Fig.~\ref{fig:Rscale} for the HT and SS membranes depicted in Fig.~\ref{fig:illust}(a) and (b), respectively.}
\label{tab:1}
\end{table}

The exponents $\nu_1$ and $\nu_2$ are typically close to unity for both HT and SS membranes, though deviations from this value are evident for membranes of zero thickness. The exponent $\nu_3$ describing the scaling of the smallest shape-tensor eigenvalue is somewhat lower than unity, as expected for a ``flat'' configuration. As for the other exponents, $\nu_3$ is also somewhat lower for $w$=0 than for $w$=0.1. Assuming that the observed scaling persists for larger membranes, the observation that $\nu_3 < \nu_1\approx\nu_2\approx 1$ suggests that the membrane is flat. In this phase, the membrane thickness grows with $L$, but at a slower rate than that of the lateral dimensions. Such behavior also characterizes self-avoiding covalent membranes, in which local excluded-volume interactions give rise to an effective bending rigidity that promotes membrane flatness.\cite{kantor1993excluded} The ``roughness'' exponent $\nu_3$ observed for linked-ring membranes here tends to be somewhat larger than that measured for covalent membranes.\cite{zhang1996molecular,gompper1997network,popova2007structure,popova2008anomalous}

\begin{figure}[!ht]
    \centering
    \includegraphics[width=\columnwidth]{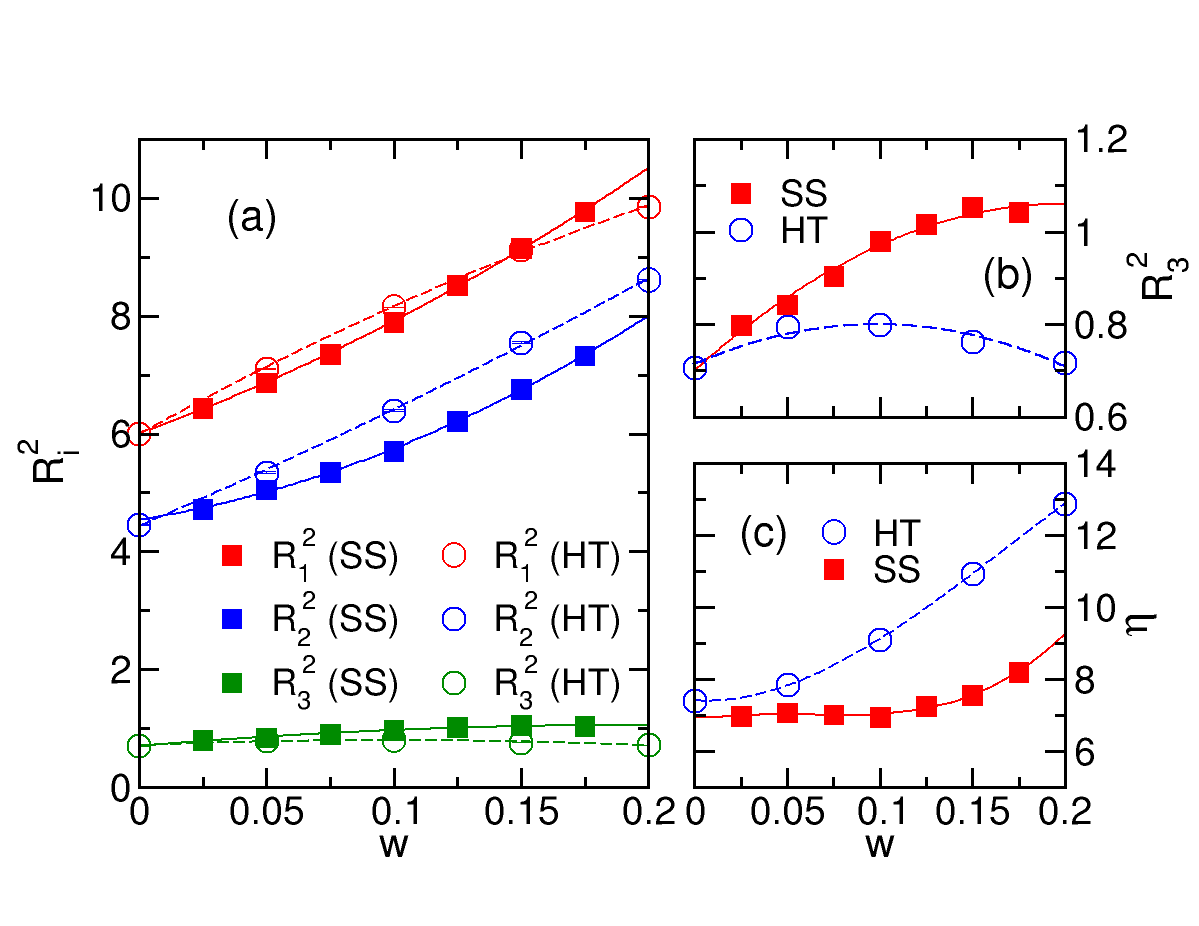}
    \caption{(a) Scaling of eigenvalues of the shape tensor as a function of ring thickness, $w$. Results are shown for a SS membrane of size $M$=10 and a HT membrane of size $M$=11. The solid and dashed curves are guides for the eye. (b) Close up of the data for $R_3^2$ from panel (a). (c) Membrane anisometry $\eta$ (defined in the text) vs ring thickness. }
    \label{fig:Rscale_width}
\end{figure}

Figure~\ref{fig:Rscale_width}(a) shows the variation of $R_i^2$ with ring thickness, $w$, for HT membranes of size $M$=11, as well as for SS membranes of size $M$=10. Both of the large eigenvalues, $R_1^2$ and $R_2^2$, increase monotonically with increasing $w$. This is indicative of an overall increase in the size of the membrane. However, as evident in the close up for $R_3^2$ in Fig~\ref{fig:Rscale_width}(b), the scaling of this eigenvalue is qualitatively different. In the case of SS membranes, this quantity increases slightly, though it appears to level off around $w\approx 0.18$. In the case of HT membranes the variation is weaker and non-monotonic, displaying a maximum near $w\approx 0.1$. Figure~\ref{fig:Rscale_width}(c) shows the membrane shape anisometry, $\eta\equiv (R_1^2+R_2^2)/2R_3^2$, with $w$.  For the SS membrane, $\eta$ is mostly constant, except at $w\geq 0.15$, where there is a small increase. By contrast, the shape anisometry of the HT membrane increases significantly over the entire range of $w$.

\begin{figure}[!ht]
    \centering
    \includegraphics[width=\columnwidth]{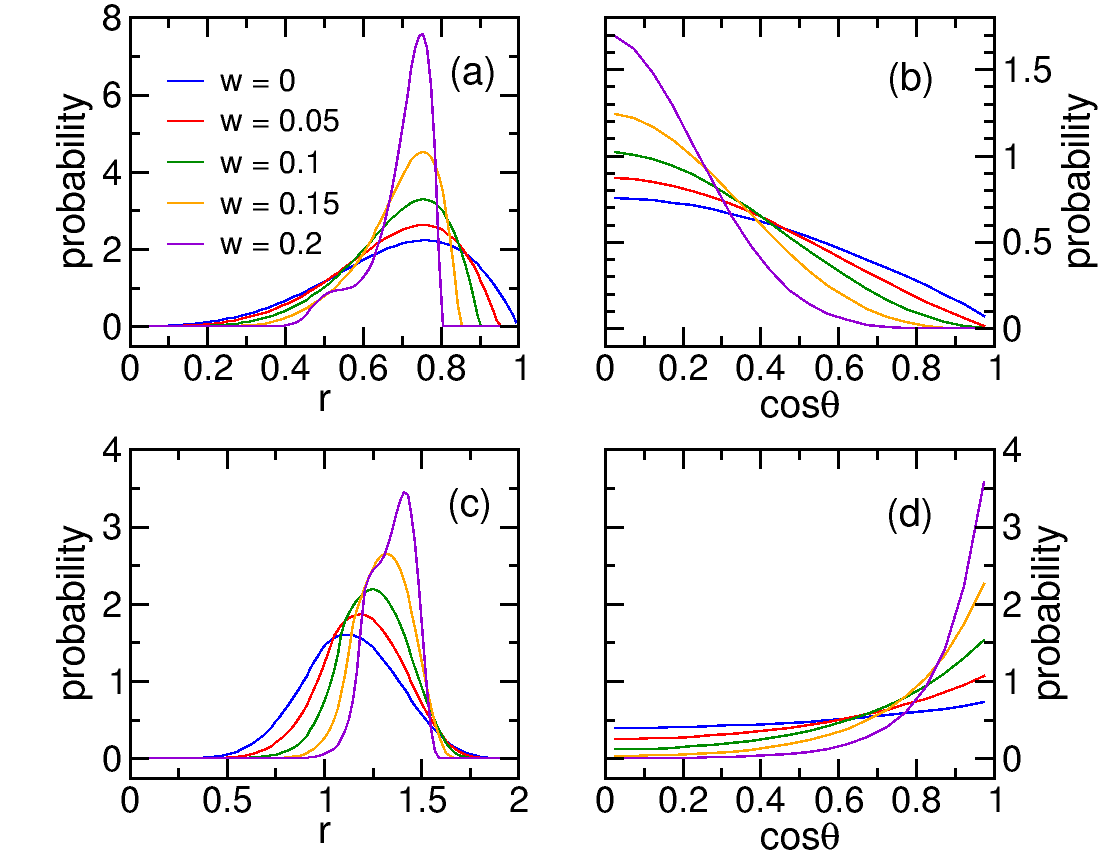}
    \caption{(a) Probability distribution for distance between centers of pairs of linked rings. Results are shown for a HT membrane of size $M$=11 for various values of $w$. (b) Distribution for $\cos\theta$, where $\theta$ is the angle between normal vectors for pairs of linked rings. The legend is same as in panel (a). (c) As in (a), except distributions for neighboring 6-valence rings. (d) As in (b), except $\theta$ is the angle between normal vectors of neighboring 6-valence rings.}
    \label{fig:pdist}
\end{figure}

Figure~\ref{fig:pdist} provides further insight into the effects of varying the ring thickness using the case of a HT membrane of size $M$=11. Figures~\ref{fig:pdist}(a) and (b) show the probability distributions for center-to-center ring distance and angle between normal vectors, respectively, for pairs of linked rings. Generally, as $w$ increases, the ring-distance distribution narrows, and the normal vectors of linked rings are increasingly likely to be perpendicular to each other (this is the relative ring orientation depicted in the illustration of Fig.~\ref{fig:illust}(d)). Figures~\ref{fig:pdist}(c) and (d) show the same two distributions except between pairs of neighboring 6-valence rings. As in Fig.~\ref{fig:pdist}(a), the distance distribution narrows, but unlike the previous case, there is an additional shift toward greater distances as $w$ increases. This key result explains the increase in $R_1^2$ and $R_2^2$ with $w$ in Fig.~\ref{fig:Rscale_width}. The sharpening of the distribution around $\theta=0$ (i.e., $\cos\theta =1$) with increasing $w$ in Fig.~\ref{fig:pdist}(d) indicates that the orientations of neighboring 6-valence rings of the HT membrane are becoming increasingly aligned as the rings become thicker, likely effecting a reduction in membrane roughness.

The results presented in Figs.~\ref{fig:Rscale}, \ref{fig:Rscale_width} and \ref{fig:pdist} {\it appear} to suggest that linked-ring membranes behave comparably to self-avoiding covalent membranes in the following manner. Independent of the membrane shape (square or hexagonal) or linking topology (square or triangular), the membrane is flat. Increasing ring thickness principally affects the pair distribution function of linked rings. This increases the mean center-to-center distance, thus increasing the membrane size quantified by $R_i^2$ in a manner analogous to increasing the tethering range of bound particles in a covalent membrane. 

\begin{figure}[!ht]
    \centering
    \vspace*{0.2in}
    \includegraphics[width=\columnwidth]{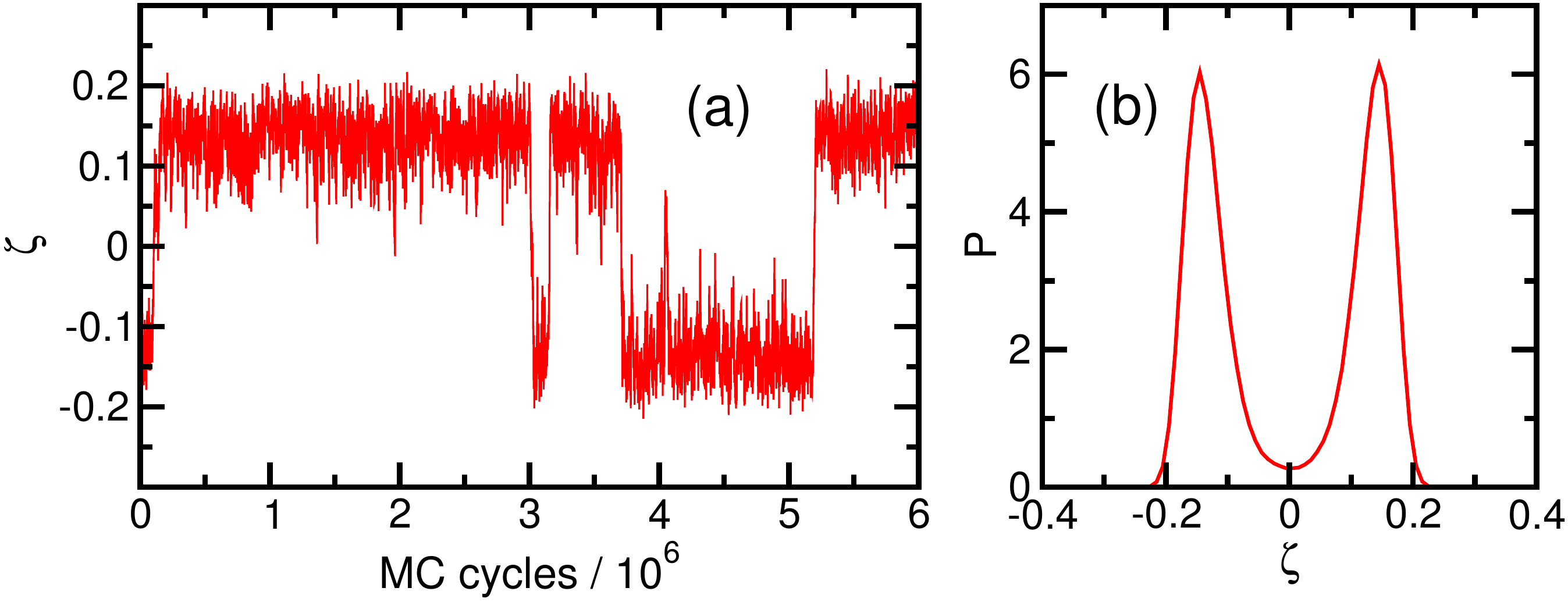}
    \includegraphics[width=0.8\columnwidth]{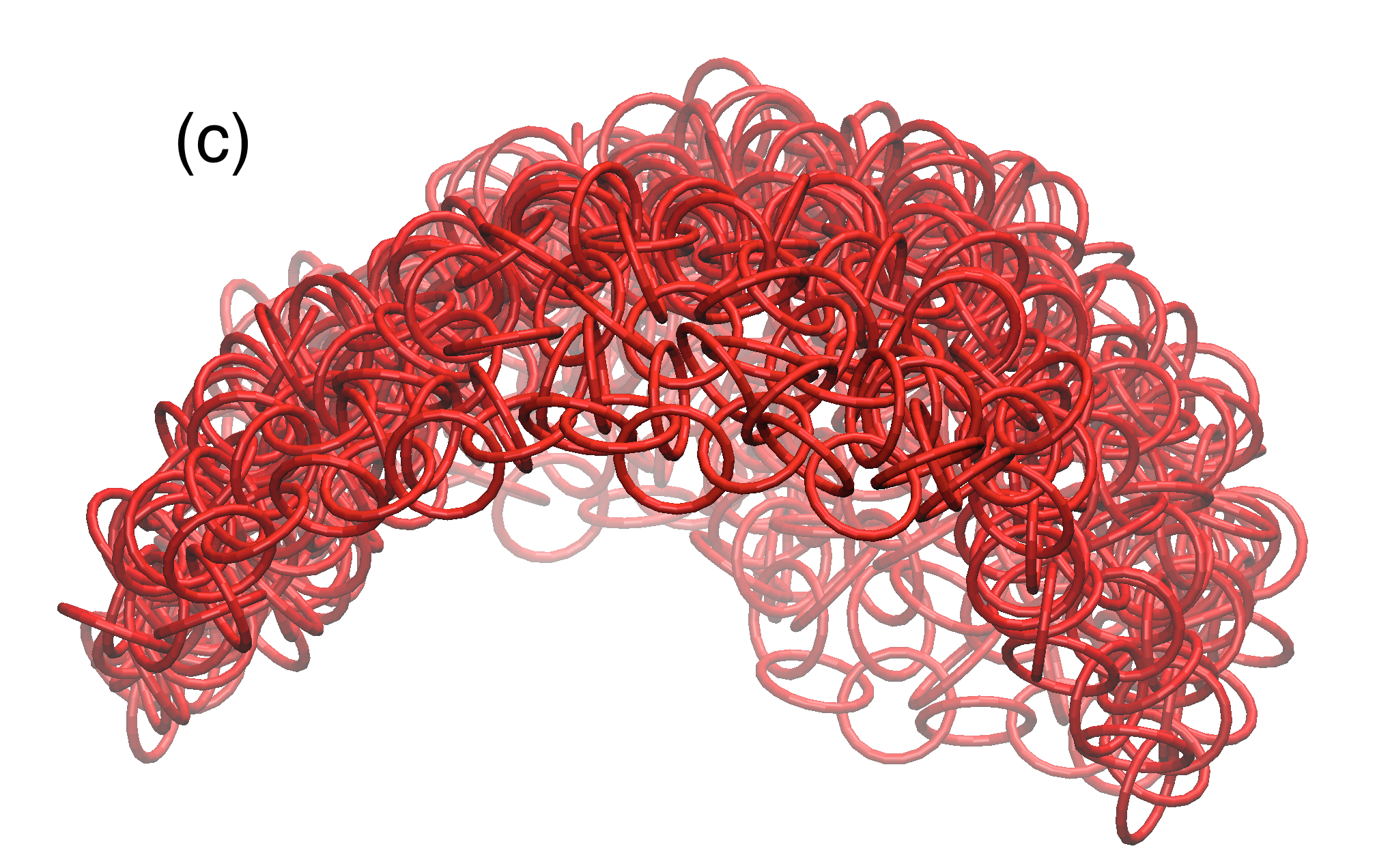}
    \caption{(a) Time dependence of $\zeta$ from a single simulation for a HT membrane of size $M$=5 and ring thickness $w$=0.15. (b) Concavity probability distribution for the same system as in (a). (c) Snapshot illustrating a HT membrane with concave shape for a system with $M$=11 and $w$=0.15.}
    \label{fig:concavity_hist}
\end{figure}

A much more interesting picture emerges, however, when we examine the membrane concavity, $\zeta$, as defined in Eq.~(\ref{eq:concavity}). Figure~\ref{fig:concavity_hist}(a) shows the time dependence of $\zeta$ for a HT membrane of size $M$=5 and ring thickness $w$=0.15. Generally, $\zeta$ tends to fluctuate about the two values of $\pm 0.13$, between which it infrequently executes rapid jumps. The corresponding probability distribution measured from an average of many such histories is shown in Fig.~\ref{fig:concavity_hist}(b). As expected, the distribution is symmetric about $\zeta=0$. In addition, it features two sharp peaks with maxima at $\pm 0.13$, whose widths are a measure of the magnitude of the fluctuations about these values. The probability at zero concavity is very low relative to the value at the maxima, consistent with the observation of infrequent and rapid transitions between the two states. Physically, this behavior corresponds to the presence of a concave membrane that periodically transitions to a new state where the concave side switches from one face to the other. Figure~\ref{fig:concavity_hist}(c) shows a snapshot of a membrane of size $M$=13 that clearly illustrates the concave shape, qualitatively similar to the concave shapes observed in microscopy studies of kinetoplasts.\cite{klotz2020equilibrium}

\begin{figure}[!ht]
    \centering
    \includegraphics[width=\columnwidth]{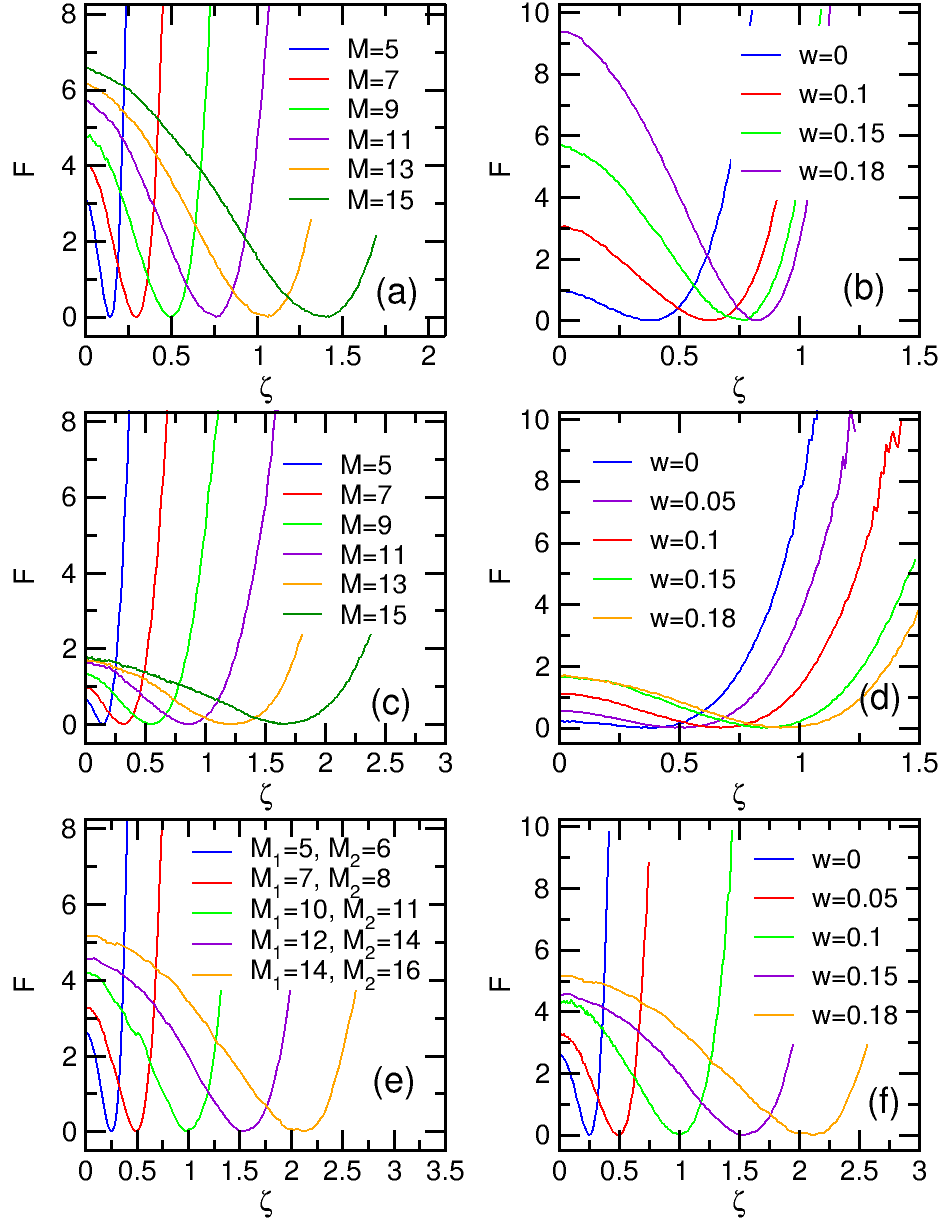}
    \caption{(a) Concavity free-energy functions for a HT membrane of ring thickness $w$=0.15. Results for various $M$ are shown. (b) As in (a), except for fixed size of $M$=11 and various $w$. (c) Free-energy functions for a SS membrane of ring thickness $w$=0.15 for various $M$. (d) As in (c) except for fixed size of $M$=11 and various $w$. (e) Free-energy functions for an ST membrane with $w$=0.15 and various $M_1$ and $M_2$. (f) As in (e), except for fixed $M_1$=10 and $M_2$=11 and various $w$.}
    \label{fig:Fzeta}
\end{figure}

Let us now examine the properties of the concavity free-energy functions, $F(\zeta)/k_{\rm B}T=-\ln {\cal P}(\zeta)$. Figure~\ref{fig:Fzeta}(a)--(f) shows a collection free-energy functions that illustrate the effects of lattice type, membrane size, and ring thickness. Note the free energy is measured in units of $k_{\rm B}T$. Since ${\cal F}(\zeta)={\cal F}(-\zeta)$, we plot functions for $\zeta\geq 0$ without any loss of information. Figure~\ref{fig:Fzeta}(a) shows functions for HT membranes of various size, each for fixed ring width of $w$=0.15. Two trends are evident. First, the most probable concavity, defined by the minimum in the free energy, $\zeta_{\rm min}$, increases with size. Second, the free energy barrier at a concavity of $\zeta=0$ increases with $M$. Thus, as the membrane size increases, it becomes increasingly unlikely that the membrane will spontaneously flip to the state with the opposite concavity. For sufficiently large size, the membrane effectively becomes locked into whichever state the system randomly selected at the beginning of the simulation. Figure~\ref{fig:Fzeta}(b) reveals that both the most probable concavity, $\zeta_{\rm min}$, and the free energy barrier height increase with increasing ring thickness. Thus, like increasing system size, increasing $w$ stabilizes the system by reducing the likelihood of concavity ``flips''.

Figures~\ref{fig:Fzeta}(c) and (d) show the effects of varying membrane size and ring thickness, respectively, for the SS membranes illustrated in Fig.~\ref{fig:illust}(b). Likewise, Figs.~\ref{fig:Fzeta}(e) and (f) show corresponding results for the ST membrane illustrated in Fig.~\ref{fig:illust}(c). The trends are mostly qualitatively consistent with those for HT membranes. In each case, there is a free-energy barrier centered at $\zeta$=0, as well as a free-energy minimum located at $\zeta_{\rm min}$, both of which tend to increase with membrane size and ring thickness. However, there are quantitative differences in the results for the HT and SS membranes. Most notably, the SS-membrane barrier height, $\Delta F\equiv F(0)-F(\zeta_{\rm min})$, is smaller than that of the HT membrane and appears to level off with membrane size and thickness. By contrast, the trends for the ST membrane are much closer to the those of the HT membrane. Specifically, $\Delta F$ increases monotonically with membrane size and ring thickness, with values considerably greater than those of SS membranes of comparable size.

\begin{figure}[!ht]
    \centering
    \includegraphics[width=\columnwidth]{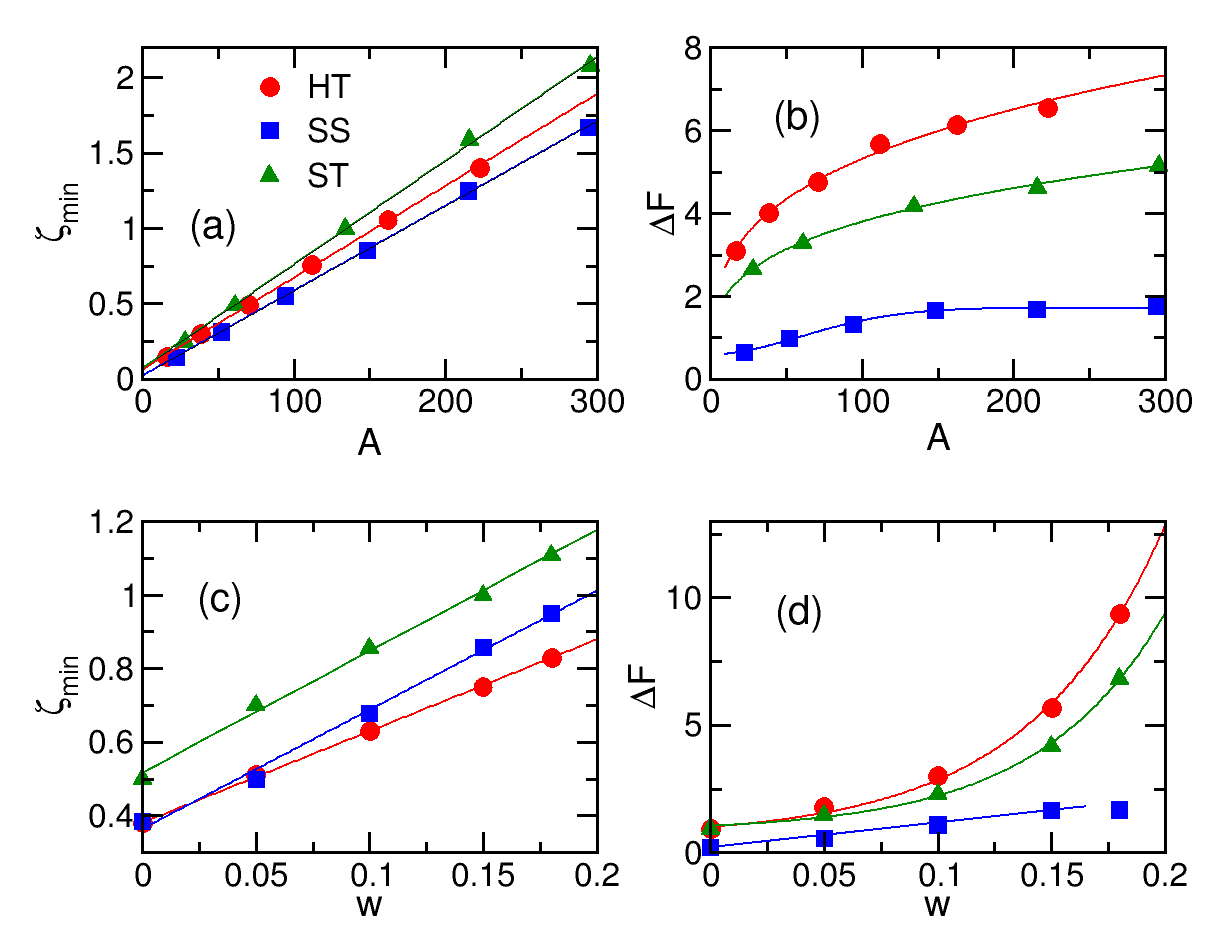}
    \caption{(a) Variation of $\zeta_{\rm min}$ with membrane area $A$, and (b) variation of the barrier height $\Delta F$ with $A$ for the three types of membrane shown in Fig.~\ref{fig:illust}. The definition of $A$ is given in the text. In each case, the ring width is $w$=0.15. (c) Variation of $\zeta_{\rm min}$ with $w$, and (d) variation of $\Delta F$ with $w$. Results are shown for the HT and SS membranes, each of size $M=11$, and for ST membranes of size $M_1$=10 and $M_2$=11.}
    \label{fig:zeta_min}
\end{figure}

Figures~\ref{fig:zeta_min}(a) and (b) show a comparison of the variation of $\zeta_{\rm min}$ and $\Delta F$ with membrane size for different membranes. As a rough measure of membrane size for HT membranes, we use the sum of the areas of the triangles formed by the 6-valence rings (see Fig.~\ref{fig:illust}(a)). The resulting membrane area is $A=c^2 l_{\rm b}^2(M-1)^2$, where $l_{\rm b}$ is the measured root-mean-square distance between neighboring 6-valence rings, and where $c^2=3\sqrt{3}/8$. A comparable measure of area for the SS membranes depicted in Fig.~\ref{fig:illust}(b) has the same form but with $c^2=1$. Likewise, for the ST membranes of Fig.~\ref{fig:illust}(c), $c^2=\sqrt{3}/2$.

Figure~\ref{fig:zeta_min}(a) shows that $\zeta_{\rm min}$ varies linearly with $A$ for each of the three membrane types. At any given $A$, the values are comparable for the different membranes. Figure~\ref{fig:zeta_min}(b) shows the monotonic increase of $\Delta F$ with $A$ for HT and ST membranes. By contrast, the much smaller barrier for SS membranes increases only for $A\lesssim 150$, following which it levels off.

Figure~\ref{fig:zeta_min}(c) and (d) shows the variation of $\zeta_{\rm min}$  and $\Delta F$, respectively, with ring thickness $w$. Results are shown for HT and SS membranes, each of size $M$=11, as well as for ST membranes of size $M_1$=10 and $M_2$=11. For each membrane, $\zeta_{\rm min}$ varies linearly with $w$. Note that the areas of the membranes at any given $w$ differ slightly, which may account for the somewhat larger values for the ST membranes. The most notable trend in Fig.~\ref{fig:zeta_min}(d) is the qualitative difference between the results for the SS membrane relative to those for the other membrane types.  The leveling off of the barrier height for $w\gtrsim$0.15 for SS membranes stands in contrast to the continuing increase for HT and ST membranes. 

A crude explanation for the linear scaling of $\zeta_{\rm min}$ with $A$ follows from employing a simple {mathematical} model in which the membrane is approximated as a small portion of a spherical surface with a uniform mass density. As described in the appendix, this model predicts an approximately linear relation between $\zeta$ and $A$, the area of the concave surface. A linear fit to the predicted curve, shown in  Fig.~\ref{fig:zeta_A_illust}, yields a slope that is comparable to, though slightly greater than the value measured for the HT membrane in Fig.~\ref{fig:zeta_min}(a) for the case of $w$=0.15. The roughness of the membrane (clearly not accounted for in the smooth surface depicted in the inset of Fig.~\ref{fig:zeta_A_illust}) likely accounts in part for the small quantitative discrepancy. 

\begin{figure}[!ht]
    \centering
    \includegraphics[width=\columnwidth]{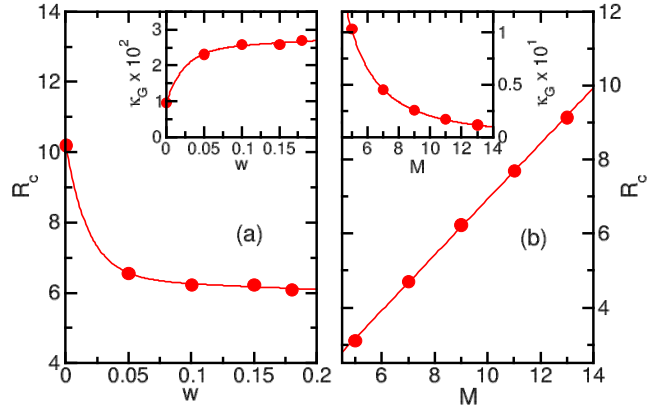}
    \caption{{(a) Variation of $R_{\rm c}$ ($\equiv 1/\sqrt{\kappa_{\rm G}}$) with ring 
                 thickness for a HT membrane of fixed size $M=9$. The inset shows the 
                 corresponding variation of the Gaussian curvature, $\kappa_{\rm G}$, with $w$. 
             (b) Variation of $R_{\rm c}$ with membrane size $M$ for a HT membrane of 
                 fixed ring thickness $w=0.15$. The inset shows the 
                 corresponding variation of the Gaussian curvature with $M$.}}
    \label{fig:gaussian}
\end{figure}

{A complementary measure of the degree of membrane concavity is the Gaussian curvature, $\kappa_{\rm G}$.
The Gaussian curvature is easily calculated at any node on a triangular mesh using the
method described by Meyer {\it et al.}\cite{meyer2003discrete} This can be conveniently
applied to an HT membrane, where each 6-valence ring is essentially a node in a triangular
mesh. We have carried out such calculations for an HT membrane and measured the mean $\kappa_{\rm G}$.
The details of the procedure are described in Section~IV of the ESI.\dag~ The insets of 
Fig.~\ref{fig:gaussian}(a) and (b) show the variation of $\kappa_{\rm G}$ with ring thickness, $w$,
and with membrane size, $M$, respectively. The main part of each panel of the figure shows the
variation of the characteristic length $R_{\rm c}\equiv 1/\sqrt{\kappa_{\rm G}}$ with $w$ and $M$. 
The results are illuminating. Perhaps the most notable point is the fact that $\kappa_{\rm G} > 0$
for all systems measured. The positive Gaussian curvature indicates that the membrane is indeed
concave, as suggested by the results for $\zeta$ above.
In Fig.~\ref{fig:gaussian}(a) we note that $\kappa_{\rm G}$ and 
$R_{\rm c}$ are only weakly dependent on the ring thickness, except at $w=0$, where the curvature 
is notably lower. By contrast, Fig.~\ref{fig:gaussian}(b) shows that the curvature rapidly decreases 
with increasing membrane size. Remarkably, $R_{\rm c}$ increases linearly with $M$. This trend
has a straightforward interpretation if the membrane is modeled as a portion of a spherical
surface as in the inset of Fig.~\ref{fig:zeta_A_illust}. In this picture $M$ is proportional to the 
diameter of the membrane as measured along its surface, and the quantity $R_{\rm c}$ is the radius
of the underlying sphere. The proportionalisty $R_{\rm c}\propto M$ then implies that the curvature
increases in a manner that fixes the angle $\theta_0$, defined in the figure. In this sense,
the membrane shape is preserved as both membrane diameter and radius of curvature co-increase.
}

Note that the presence of intrinsic curvature complicates the simple interpretation, based on the scaling of $R_i^2$ in Fig.~\ref{fig:Rscale}, that the membrane is flat, i.e. that the lateral dimensions of the membrane as measured along the surface grow faster than its thickness with increasing system size. As a result of this curvature the exponents $\nu_1$ and $\nu_2$ are expected to be somewhat lower than unity, while the roughness exponent, $\nu_3$, is expected to be larger than a value determined solely by out-of-plane fluctuations in the membrane shape. To estimate the magnitude of these effects we employ again the spherical-surface model described above. As shown in the appendix, this simple model predicts minimal effect on the values of $\nu_1$ and $\nu_2$. In addition, it suggests that the contribution to $R_3^2$ is small compared to that from the effects of membrane roughness. Consequently, it is reasonable to interpret the scaling results as implying linked-ring membranes are flat.

The trends evident in Figs.~\ref{fig:Fzeta} and \ref{fig:zeta_min} suggest that concavity appears to be an intrinsic property of the simple  membranes composed of rigid interlocking circular rings examined here. The growth of the free-energy barrier $\Delta F$ with membrane area for membranes with triangular linking topology (HT and ST membranes) suggests that a concave configuration is increasingly preferred as membranes of this type grow in size. The situation is qualitatively different, however, for membranes with a square linking topology (SS membranes). In this case the barrier is much smaller and approaches a constant value as $A$ increases, and thus, the tendency toward concavity is much weaker. The similarity of the behavior for HT and ST membranes suggests the qualitatively different results for the SS membrane is not due to the shape of the membrane (square, rather than hexagonal), but rather its linking topology (triangular, rather than square). It seems that the ``tighter'' triangular linking topology (i.e. a higher linking valence) underlies the spontaneous formation of stable concave structures. Increasing the ring thickness has the combined effect of swelling the size of the membrane, as noted in Fig.~\ref{fig:Rscale_width}, and increasing the tendency toward concave shapes, as noted in Fig.~\ref{fig:zeta_min}(d).  

\begin{figure}[!ht]
    \centering
    \includegraphics[width=\columnwidth]{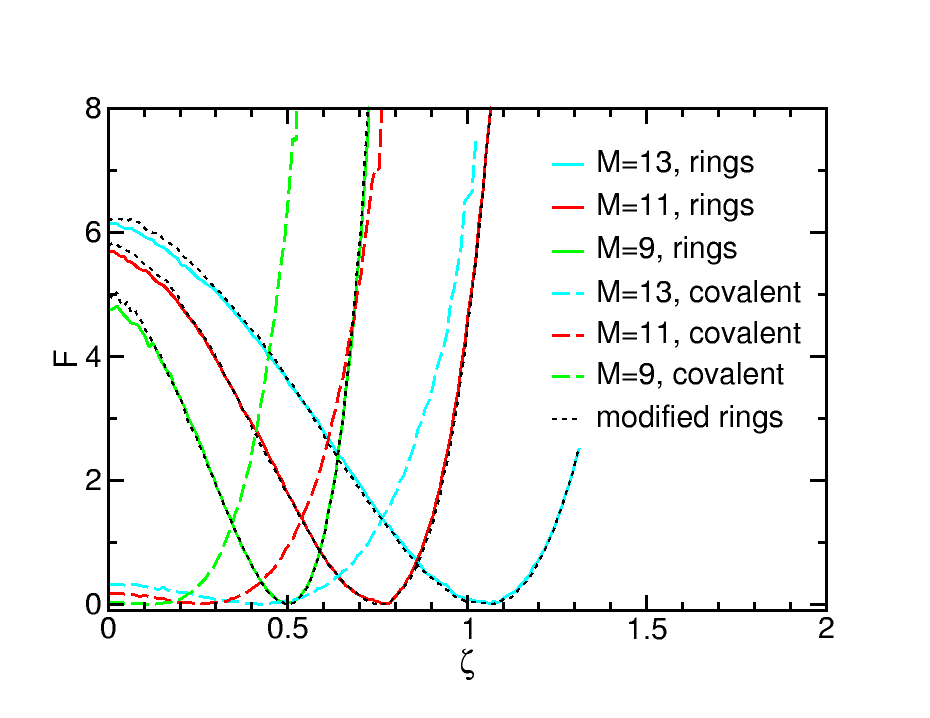}
    \caption{Concavity free-energy functions for hexagonal-shaped membranes. Results for various sizes are shown for a HT membrane with ring thickness $w$=0.15 and for an equivalent covalent membrane, as defined in the text. The dashed black curves show results for modified HT membranes where excluded-volume and linking interactions were considered only for pairs of rings that share a linking partner.}
    \label{fig:Fzeta_covalent}
\end{figure}

It is instructive to compare the concavity of the linked-ring membranes to that of  covalent membranes of a type examined in previous studies. As an example, consider a membrane composed of hard spherical particles of diameter $\sigma$ connected through a fixed network of tethers, each to a small number of neighboring particles. The interaction energy between tethered particles is zero, unless the particles overlap or distance between them exceeds some limit, $b$, in which case it is infinite. Such self-avoiding athermal membranes are known to be flat. Using this model we have carried out simulations of hexagonal membranes with the triangular tethering network. This is analogous to the HT membrane shown in Fig.~\ref{fig:illust}(a), with the 6-valence rings replaced by hard spheres, and the 2-valence rings effectively replaced by the tethers. To make the comparison meaningful, we choose the parameter values of $\sigma$ and $b$ to yield the mean and variance of the distance between tethered particles to match that between the centers of the 6-valence rings in the linked-ring membrane. Figure~\ref{fig:Fzeta_covalent} shows concavity free-energy functions for linked-ring membranes of ring thickness $w$=0.15 and the corresponding covalent membranes for various $M$. As expected, $F(\zeta)$ for the covalent membrane shows a negligible barrier for any membrane size, in stark contrast to the linked-ring membranes. As well, functions rise steeply at much smaller $\zeta$ than those for covalent membranes. 

The results shown in Fig.~\ref{fig:Fzeta_covalent}  highlight the fact that the spontaneous concavity of small linked-ring membranes does not arise simply from the presence of excluded-volume interactions, which are also present for the covalent membrane. Instead, it seems to be a consequence of the form of the anisotropy in these interactions; that is, the excluded volume depends strongly on the relative orientation of any pair non-linked rings, as does the range of the accessible center-to-center separation distance between linked rings. A loose analogy is the spontaneous entropy-driven orientational ordering in colloidal liquid crystals arising from the orientational anisotropy in the excluded-volume interaction between elongated colloidal particles.\cite{vroege1992phase} 

As noted in the introduction, self-avoiding covalent membranes are flat as a result of {\it local} excluded-volume interactions, which give rise to an effective bending rigidity, while interaction sites separated by longer distances (as measured along the membrane) play no significant role.\cite{kantor1993excluded} To determine whether the concavity of linked-ring membranes likewise arises from {\it local} anisotropic excluded-volume interactions, we carry out a calculation for a linked-ring system modified as follows. For pairs of rings that are either linked or which have a common linking partner, the tests for linking and overlap are implemented as before. For all other pairs of rings, the tests for linking and overlap are ignored in the MC algorithm. This means that pairs of rings separated by two or more links are permitted to overlap, for example in a ``taco'' configuration. The simulations were carried out for HT membranes of various sizes for $w$=0.15. The calculated free-energy functions are overlaid shown as dashed lines in Fig.~\ref{fig:Fzeta_covalent}. The functions for the modified system are virtually identical to those for the original membranes, with only a small increase in $F$ for low $\zeta$. We conclude that membrane concavity does indeed arise from {\it local} interactions between rings.
 
As noted earlier, a key motivation for the present study is to provide some insight into the observed properties of kinetoplasts, structures consisting of thousands of interlinking circular DNA molecules. Two simulation results stand out as possibly relevant for kinetoplasts. The first is the monotonic increase in the membrane size with increasing ring thickness, evident in Figs.~\ref{fig:Rscale} and \ref{fig:Rscale_width}. By comparison, Soh {\it et al.} found that kinetoplasts increased in size with the effective width of the DNA rings, which is controlled by varying the ionic strength of the solvent.\cite{soh2020ionic} The second result is the emergence of intrinsic curvature giving rise to the concave structures, as seen for example in the snapshot of Fig.~\ref{fig:concavity_hist}, which is also a property of kinetoplasts. The other naturally occurring planar catenated network, the capsid of the HK97 virus, is also found in a strongly curved state.\cite{zhou2015protein} Although curved surfaces can easily be constructed from flat surfaces through various means, such as a tailor-like procedure of cutting and removing wedges\cite{grosberg2020human} or through a purse-string mechanism, it is surprising and notable that the simple linking networks such as those shown in Fig.~\ref{fig:illust} exhibit this property.

Some caveats are in order here. To manage the computational cost our simulation model is necessarily simplistic and ignores numerous features of the real system whose effects may well be non-negligible. For example, the kinetoplast DNA mini-circles have a typical contour length of several Kuhn lengths and are thus flexible objects, in contrast to the rigid circular rings in the model. In addition, the membrane linking structures shown in Fig.~\ref{fig:illust} are somewhat arbitrary, though convenient, choices. An alternative and perhaps more realistic approach might be selection of a random linking topology chosen in a manner as in Ref.~\citen{diao2015orientation}. Finally, we note the relatively small size of the membranes employed in the simulation. Even the largest model membranes contain far fewer rings than kinetoplasts. They are also far fewer than the typical number of nodes or particles used in simulations of covalent membranes. Unfortunately, simulation of larger catenated networks is not computationally feasible at present. Consequently, it remains an open question as to whether the observed trends will persist for much larger membranes. Still, future simulations may eventually remedy these limitations by using more realistic models, and we  view the present simulation study as an important first step toward understanding the behavior of catenated networks such as kinetoplasts. 

\section{Conclusions}

In this study, we use Monte Carlo simulations to examine the statistical properties of ``membranes'' composed of 2D networks of linked rings. This work is largely inspired by recent experiments studying the physical properties of kinetoplasts,  chain-mail-like structures found in the mitochondria of trypanosome parasites consisting of thousands of catenated circular DNA molecules. To keep the simulations computationally feasible, we employ a highly simplified model using hard, rigid, circular rings that are linked together in a regular lattice pattern, and consider membranes that are effectively much smaller than kinetoplasts. Generally, the scaling of the average membrane dimensions with system size suggest that the networks are flat, in the sense that the lateral dimensions grow much faster than the membrane thickness. Increasing ring thickness tends to swell the membrane, qualitatively consistent with observations from kinetoplast experiments. Remarkably, we find that the membranes tend to form concave structures that qualitatively resemble the shapes observed in kinetoplasts. This feature is of entropic origin and arises from {\it local} anisotropic excluded-volume interactions between rings. The degree of concavity increases with ring thickness and tends to be more pronounced in networks with a higher linking valence. 

Future work will focus on refining the model to make it better resemble the experimental system. Two relevant features to incorporate are flexibility of the rings and a random linking topology with the ``correct'' mean valence. It will be of interest to determine whether the observed membrane concavity is affected by such changes. Another topic of interest is the effect of holes on the conformational properties of linked-ring membranes, a feature recently studied in the context of covalent membranes.\cite{yllanes2017thermal} A longer-term goal is developing a more computationally efficient coarse-grained membrane model using the measured properties of the present model system. Such a model would effectively facilitate simulations of much larger membranes that better resemble kinetoplasts.

\appendix

\section{Appendix}

In this appendix, we examine a simple model of a concave surface to help understand the observed variation of $\zeta_{\rm min}$ with system size in Fig.~\ref{fig:Fzeta}(a) and \ref{fig:zeta_min}(a). The model is also used to estimate the effects of intrinsic curvature on the scaling exponents obtained from fits to the data in Fig.~\ref{fig:Rscale}.

\begin{figure}[!ht]
    \centering
    \includegraphics[width=\columnwidth]{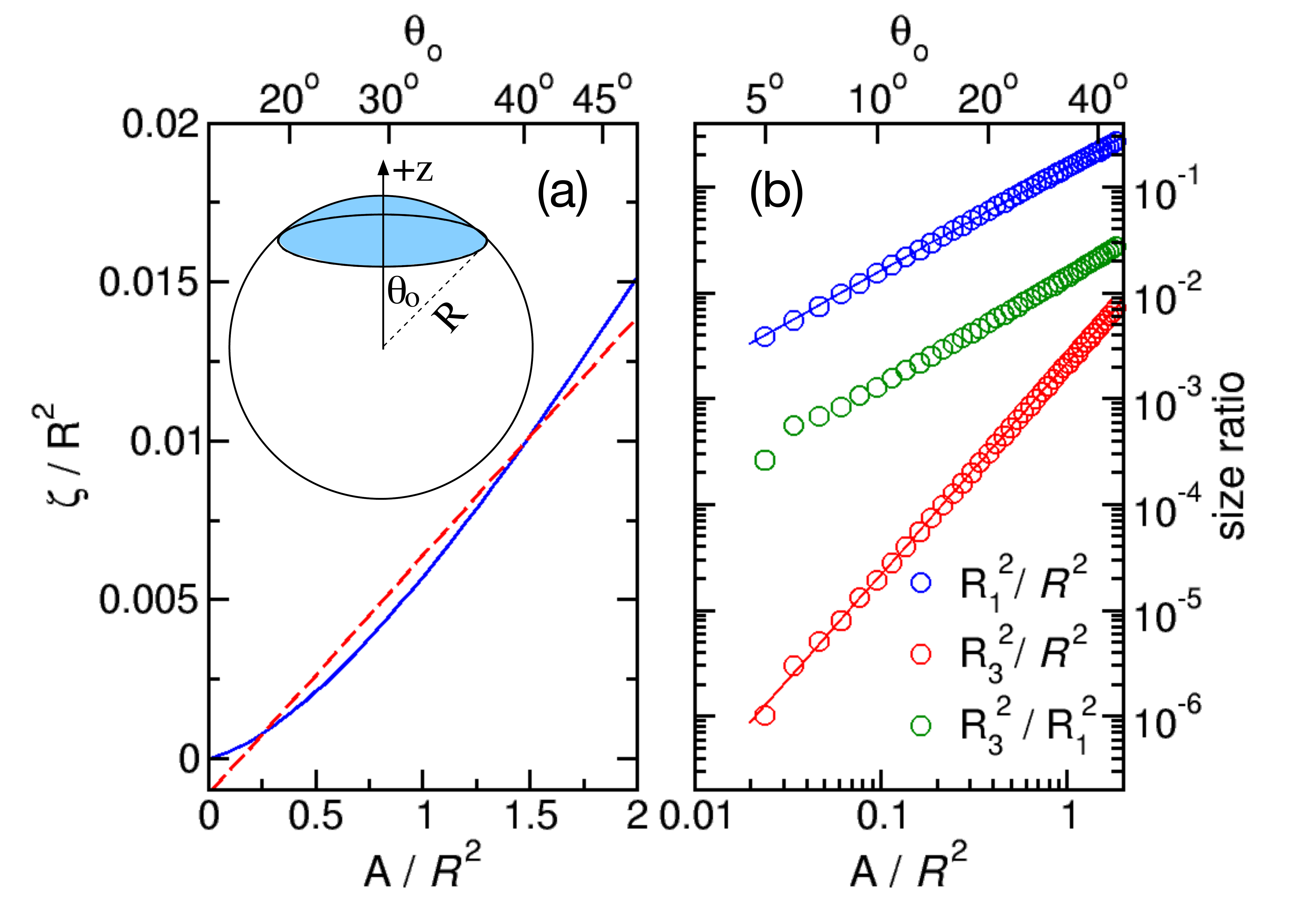}
    \caption{(a) Variation of scaled concavity, $\zeta/R^2$, with scaled surface area, $A/R^2$, for a surface constructed from a portion of a sphere of radius $R$. The dashed red line is a linear fit to the data. The inset shows an illustration of the surface (shaded in blue) in relation to the underlying spherical surface and also shows the definition of $\theta_0$. (b) Scaled shape-tensor eigenvalues, $R_1^2/R^2$ and $R_3^2/R^2$ vs scaled surface area $A/R^2$ for the spherical shell model shown in the inset of Fig.~\ref{fig:zeta_A_illust}. The solid curves show power-law fits to the two data sets, which yield effective scaling exponents of $\nu_1$=0.977 and $\nu_3$=1.997. The ratio $R_3^2/R_1^2$ is also shown. }
    \label{fig:zeta_A_illust}
\end{figure}

Consider a two-dimensional surface constructed from a portion of the surface of a sphere of radius $R$, as illustrated in the inset of Fig.~\ref{fig:zeta_A_illust}. The surface is of uniform mass density. It can be shown that the area of the surface is 
\begin{eqnarray}
A=2\pi R^2(1-\cos\theta_0),
\label{eq:A}
\end{eqnarray}
and that the $z$-component of the center of mass is given by $z_{\rm cm}=\frac{1}{2}R(1-\cos\theta_0)$, where the angle $\theta_0$ is defined in the figure. The concavity $\zeta$ can be calculated as
\begin{eqnarray*}
\zeta = \frac{1}{A} \int_{S} \sqrt{x^2+y^2} (z_{\rm cm}-z) da.
\end{eqnarray*}
Using spherical coordinates to evaluate the integral, it is easily shown that
\begin{eqnarray}
\frac{\zeta}{R^2} = \frac{\left[{\textstyle\frac{1}{3}}\sin^3\theta_0 -{\textstyle\frac{1}{4}}(1+\cos\theta_0)(\theta_0-\sin\theta_0\cos\theta_0)\right]}{1-\cos\theta_0}.
\label{eq:zeta_theory}
\end{eqnarray}
In Fig.~\ref{fig:zeta_A_illust}(a), the blue curve shows the variation of $\zeta/R^2$ vs $A/R^2$, calculated using Eqs.~(\ref{eq:A}) and (\ref{eq:zeta_theory}), respectively, for $\theta_0$ ranging from 0 to 45$^\circ$. The function is only slightly curved across this range. A linear fit to the data, shown as the dashed red line, yields a slope of $m=0.0075$, and so $\zeta/R^2\approx 0.0075 A/R^2$.

In applying this model to analyze the concavity of linked-ring membranes, we note that the curvature radius $R$ may itself depend on the membrane size, as quantified by the total ring number $N$, as do both $A$ and $\zeta$.  Nevertheless, the fact that $\zeta(N)/R(N)^2\propto A(N)/R(N)^2$ implies $\zeta(N) \propto A(N)$ with the same proportionality constant, in this case the value $m=0.0075$. In Fig.~\ref{fig:zeta_min}(a) the most probable concavity, $\zeta_{\rm min}$ follows the approximate scaling $\zeta_{\rm min} \approx 0.0061A$ for the case of HT membranes with a triangular bonding linking network. The proportionality constant of 0.0061 compares reasonably well to the value of 0.0075 obtained from the linear fit to the curve in Fig.~\ref{fig:zeta_A_illust} above. The difference is attributable to several approximations, including the choice of shape and curvature of the model surface and the neglect of thermal fluctuations in the surface. Indeed, the presence of undulations in the surface would contract its effective area (when projected onto a sphere) for a given $M$, contributing to the observed smaller value of the proportionality constant. Although the model neglects such subtle effects, it nevertheless provides a simple explanation for the variation of the most probable concavity with membrane size. Finally, note that this model must eventually break down for sufficiently large $A/R^2$, most obviously in the regime $\theta_0>\pi/2$, where the predicted membrane edge length decreases with membrane area. 

The model can also be used to estimate the effects of concavity on the scaling properties of the shape-tensor eigenvalues, $R_i^2$. In Fig.~\ref{fig:Rscale}, we noted that the smallest eigenvalue $R_3^2$ is typically an order of magnitude smaller than $R_1^2$ and $R_2^2$. In addition, the corresponding scaling exponents $\nu_1$ and $\nu_2$ are close to unity while $\nu_3$ are somewhat smaller. Taken together, these observations appear to suggest that the membrane is flat, in the sense that the lateral size grow faster than the (smaller) membrane thickness. However, long-range curvature complicates this interpretation. The concavity leads to an additional contribution to $R_3^2$, which could grow with membrane size. This is expected to increase the effective exponent, $\nu_3$, beyond the value arising solely from membrane shape fluctuations, as well as reduce $\nu_1$ and $\nu_2$ from unity. 

To quantify these effects, we again employ the smooth concave surface illustrated in the inset of Fig.~\ref{fig:zeta_A_illust}(a) and calculate the variation of the scaled eigenvalues $R_1^2/R^2$ and $R_3^2/R^2$ with scaled surface area, $A/R^2$. (Note that $R_1^2=R_2^2$, by symmetry.) The results are shown in Fig.~\ref{fig:zeta_A_illust}(b) for $\theta_0\leq 45^\circ$. In this range, the eigenvalues both approximately exhibit power-law scaling. The effective scaling exponent for $R_1^2$ is $\nu_1=0.977$, very close to the value of unity expected for flat membranes without long-range curvature. The effective exponent for $R_3^2$ of $\nu_3=1.997$ is considerably larger. However, note that the ratio of the eigenvalues, $R_3^2/R_1^2$, is very small, varying between $3\times 10^{-4}$ and $3\times 10^{-2}$. By contrast, $R_3^2/R_1^2={\cal O}(10^{-1})$ in Fig.~\ref{fig:Rscale}. This suggests that the value of $R_3^2$ is mainly determined by the shape fluctuations in the linked-ring membranes rather than by long-range curvature. We note that the exponent $\nu_3$ extracted from fits to the data in Fig.~\ref{fig:Rscale} and listed in Table~\ref{tab:1} tend to be somewhat larger than values (typically $\approx 0.6$) measured for covalent membranes. It is possible that membrane curvature, which leads to such a steep increase in $R_3^2$ with area in Fig.~\ref{fig:zeta_A_illust}(b), contributes to this larger value. Regardless, the combined results of Fig.~\ref{fig:Rscale} and \ref{fig:zeta_A_illust}(b) suggest that the lateral dimension of a curved linked-ring membrane (measured along the surface) grows faster than its thickness as the total ring number $N$ increases.

\section*{Author Contributions}
JP wrote the simulation code, carried out some of the simulations and data analysis, and wrote part of the article. EG carried out some of the simulations and data analysis. AK wrote part of the article, provided experimental insight, and oversaw the project.

\section*{Conflicts of interest}
There are no conflicts to declare.

\section*{Acknowledgements}
JP acknowledges funding from the Natural Sciences and Engineering Research Council of Canada (NSERC) and is grateful to Compute Canada for use of their computational resources. This material is based upon work supported by the National Science Foundation under Grant No. 2105113.


%

\end{document}